\documentstyle[11pt]{article}
 
\begin{document}

\baselineskip=14pt plus 0.2pt minus 0.2pt
\lineskip=14pt plus 0.2pt minus 0.2pt

\newcommand{\be}{\begin{equation}}
\newcommand{\ee}{\end{equation}}
\newcommand{\da}{\dagger}
\newcommand{\dg}[1]{\mbox{${#1}^{\dagger}$}}
\newcommand{\hlf}{\mbox{$1\over2$}}
\newcommand{\lfrac}[2]{\mbox{${#1}\over{#2}$}}
\newcommand{\nsz}[1]{\mbox{\normalsize ${#1}$}}

\begin{flushright}
quant-ph/9811075 \\
LA-UR-98-727 \\
\end{flushright} 

\begin{center}
\Large{\bf Time-dependent Schr\"odinger equations \\
having isomorphic symmetry algebras.\\  
I. Classes of interrelated equations. \\}
 
\vspace{0.25in}

\large
\bigskip

Michael Martin Nieto\footnote{\noindent  Email:  
mmn@lanl.gov}\\
{\it Theoretical Division (MS-B285), Los Alamos National Laboratory\\
University of California\\
Los Alamos, New Mexico 87545, U.S.A. \\}
 
\vspace{0.25in}

 D. Rodney Truax\footnote{Email:  truax@ucalgary.ca}\\
{\it Department of Chemistry\\
 University of Calgary\\
Calgary, Alberta T2N 1N4, Canada\\}
 
\normalsize

\vskip 20pt
\today

\vspace{0.3in}

{ABSTRACT}
 
\end{center}
\begin{quotation}
\baselineskip=.33in
In this paper, we focus on a general class of Schr\"odinger equations 
that are time-dependent and quadratic in $X$ and $P$.  We transform 
Schr\"odinger equations in this class, 
via a class of time-dependent mass equations, to a class of solvable 
time-dependent oscillator equations.  This transformation consists of 
a unitary transformation and a change in the ``time'' variable.  We 
derive mathematical constraints for the transformation and 
introduce two examples.
\vspace{0.25in}


\end{quotation}

\newpage

\baselineskip=.33in

\section{ Introduction}
Finding analytical solutions to time-dependent Schr\"odinger 
equations has been a mathematical problem of considerable interest.  
Such equations are relevant to the study of dissipative systems in 
quantum theory. Solutions to the time-dependent Schr\"odinger equation 
(setting $\hbar=m=1$)
\begin{equation}
\left\{H-i\partial_t\right\}\Psi(x,t)=0,\label{e:int1}
\end{equation}
where the Hamiltonian $H$ is time-dependent, describe the evolution 
of such systems.

Several calculations for a general class of Hamiltonians that are quadratic 
in $x$ and $p$ have been studied {\cite{mk1}}-{\cite{ck1}}.  
We write the following general form for these Hamiltonians
\begin{equation}
H_1=[1+k(t)]{{p^2}\over{2}}+\lfrac{1}{2}h(t)\left(xp+px\right)+g(t)p
+h^{(2)}(t)x^2+h^{(1)}(t)x+h^{(0)}(t),\label{e:int12} 
\end{equation}
where $k$, $h$, $g$, and $h^{(j)}$, $j=0,1,2$, are suitably well-behaved 
real functions of time.  We designate such time-dependent Hamiltonians 
by $TQ$.

A subclass of time-dependent Hamiltonians, $TM$, with 
``time-dependent masses'' is {\cite{ek1}}-{\cite{spk1}},   
\begin{equation}
H_2=f(t){{p^2}\over{2}}+f^{(2)}(t)x^2,\label{e:int8}
\end{equation}
where $f$ and $f^{(2)}$ are suitably well-behaved real functions of time.
Using  Lewis invariants {\cite{hrl1,lr1}}, analytical 
solutions have been obtained for some Schr\"odinger equations with this 
type of 
Hamiltonian {\cite{ca2,fmf1}}.  Often, the function $f(t)$ 
has the form $\exp{(\pm\Upsilon t)}$,  $\Upsilon$ a real constant.  

A second subclass of time-dependent Hamiltonians, $TO$, is 
the time-dependent harmonic oscillator in one dimension.  This has 
been studied extensively {\cite{hrl1}}-{\cite{nt2}}.  Its Hamiltonian is
\begin{equation}
H_3={{p^2}\over{2}}+g^{(2)}(t)x^2+g^{(1)}(t)x+g^{(0)}(t),\label{e:int4}
\end{equation}
where  
the coefficients $g^{(j)}(t)$, $j=1,2,3$, are suitably well-behaved 
real functions of time.  One of the earliest symmetry techniques used 
to solve this equation was the method of Lewis invariants 
{\cite{hrl1,lr1}}.  Its Lie space-time symmetry algebra has 
been identified as $sl(2,{\bf R})\Box w_1$ by one of the authors 
{\cite{drt1}} and its complexification, 
$su(1,1)\Box w_1^c$, has been used to construct solution spaces 
for Eq. (\ref{e:int1}) \cite{drt2}, \cite{gt1}-\cite{nt2}.  
The subalgebra $w_1$ is a Heisenberg-Weyl algebra in one dimension and 
$w_1^c$ is its complexification.

In this paper
we show that the three classes of Schr\"odinger equations  
mentioned above can be interrelated by transformations.  
Generally speaking, these  
transformations can be given by two actions: a unitary 
transformation ($TQ\rightarrow TM$) and a change in ``time'' variable 
($TM\rightarrow TO$).  After giving notation for the three classes 
of Schr\"odinger equations in Section 2, 
we describe the unitary transformation and the change in ``time'' 
variable in Section 3.   In Section 4, we apply our analysis to 
two TM systems commonly found in the literature. (For example, see Refs. 
{\cite{lls1,spk1}}.)  We close with a short summary.

In paper II \cite{II} we  continue by studying an algebraic 
approach to solving Schr\"odinger equations for all three classes of 
systems, TQ, TM, and TO.  We shall show that these three systems have 
isomorphic Schr\"odinger algebras, 6-dimensional Lie algebras of space-time 
symmetries.  A subalgebra, having the 
structure of an oscillator algebra, will be used to derive expressions for 
number-state, coherent-state, and squeezed-state wave functions for each of 
the three classes of systems.  
Expectation values and uncertainty products will also be obtained and the 
classical equations of motion determined.  

Elsewhere \cite{else} we will concentrate on the  $TM$ systems discussed 
in Section 4 of this paper. We will apply the general 
procedures worked out here and in paper II to obtain number states, 
coherent states, and squeezed states for the $TM$ systems and examples of 
symmetry-related $TO$ and $TQ$ systems.  Our treatments of  $TM$ 
systems will be detailed and new results will emerge.  


\section{Notation}

For computational purposes, it is more convenient to rearrange the 
Schr\"odinger equation (\ref{e:int1}), with quadratic Hamiltonian 
(\ref{e:int12}), into the form 
\begin{eqnarray}
S_1\Phi(x,t) & = & \left\{[1+k(t)]\partial_{xx}+2i\partial_{t} 
+h(t)\left(-ix\partial_x-i/2\right)+g(t)(-i\partial_{x})\right. 
\nonumber\\
 &   & \hspace{2cm} \left.-2h^{(2)}(t)x^2-2h^{(1)}(t)x-2h^{(0)}
(t)\right\}\Phi(x,t) =  0,\label{e:not1}
\end{eqnarray}
where $k$, $h$, $g$, and $h^{(j)}$, $j=0,1,2$ are suitably well-behaved, 
real  functions of $t$.  

Next, we introduce the following operator algebra
\begin{eqnarray}
 & T=i\partial_{t},~~~~~P=-i\partial_x,~~~~~X=x,~~~~~I=1, & 
\label{e:not8a}\\*[1mm]
 & P^2=-\partial_{xx},~~~~~X^2=x^2,~~~~~D=\lfrac{1}{2}\left(XP+PX\right) 
= -ix\partial_x-i/2. & \label{e:not8b}
\end{eqnarray}
These operators have the following nonzero commutation relations
\begin{eqnarray}
&[X,P] = iI,&   \label{e:not12a}  \\
&[X^2,P^2]=4iD,~~~[D,X^2]=-2iX^2,~~~[D,P^2]=2iP^2,&   \label{e:not12b} \\
& [P^2,X]=-2iP,~~~[X^2,P]=2iX, ~~~
  [D,X]=-iX,~~~[D,P]=iP. & \label{e:not12c}
\end{eqnarray}
This Lie algebra of operators has a structure isomorphic to 
$sl(2,{\bf R})\Box w_1$, a Schr\"odinger algebra.  The 
operators $\{X,P,I\}$ form  
a basis for a Heisenberg-Weyl algebra, $w_1$ (see Eq. (\ref{e:not12a})), 
and the  operators $\{X^2,P^2,D\}$ form a basis for the special 
linear algebra $sl(2,{\bf R})$ (see Eq. (\ref{e:not12b})).

When we express the Schr\"odinger equation ({\ref{e:not1}}) in terms 
of these operators, we obtain 
\begin{eqnarray}
{S}_1\Phi(x,t) & = & \left\{-[1+k(t)]P^2+2T+h(t)D+g(t)P\right.
\nonumber\\
  &   & \hspace{1cm} \left.-2h^{(2)}(t)X^2-2h^{(1)}(t)X-
2h^{(0)}(t)I\right\}\Phi(x,t) =  0.\label{e:not16a}  
\end{eqnarray}
Equations of the type (\ref{e:not16a}) 
are time-dependent quadratic Schr\"odinger equations, the class $TQ$.

Next, we turn to those Schr\"odinger equations (\ref{e:int1}) which 
have Hamiltonians (\ref{e:int8}).  With the operator 
notation, (\ref{e:not8a}) and (\ref{e:not8b}), we write the 
Schr\"odinger equation as
\begin{eqnarray} 
S_2\Theta(x,t) & = & \left\{-f(t)P^2+2T
-2f^{(2)}(t)X^2-2f^{(1)}(t)X
-2f^{(0)}(t)I\right\}\Theta(x,t) =  0. \label{e:not16b}
\end{eqnarray} 
Eq. (\ref{e:not16b}) is representative of the so-called ``time-dependent 
mass'' equations.  We have denoted this class of by $TM$.  [The term 
``time-dependent mass" comes from the fact that $f(t)$ multiplies 
$P^2$, just as $1/m$ would if we did not have units $m=1$.] 
 
The second class of Schr\"odinger equations, the time-dependent
oscillator equations, denoted by $TO$, can be written as 
(here the ``time'' variable is indicated with a prime)
\begin{equation}
{S}_3\Psi(x,t') = \left\{-P^2+2T'-2{g}^{(2)}
(t')X^2 -2{g}^{(1)}(t')X-2{g}^{(0)}(t')I\right\}\Psi
(x,t') = 0.\label{e:not16c}
\end{equation}  
In the next section, we derive a connection between $t$ and $t'$.

We emphasize again that the two classes of equations, $TM$ and $TO$, 
are subclasses of the class $TQ$.  Here, our main focus will be 
on answering the question: Do transformations exist that relate a given 
Schr\"odinger equation in one class to a Schr\"odinger equation in another 
class?  In the next section, we will find a unitary transformation 
that interrelates all $TQ$ Schr\"odinger equations 
(\ref{e:not16a}).  Then, we identify the conditions that allow a $TQ$ 
equation to be transformed into a $TM$ equation 
(\ref{e:not16b}).  That equation can in turn be mapped  
into a solvable Schr\"odinger 
equation (\ref{e:not16c}) in the class $TO$.  Also, we derive conditions 
for the inverse transformations.  That such  
transformations exist is due to  the classes 
$TM$ and $TO$ being subclasses of $TQ$.


\section{The Transformation}

\subsection{The Form of the Unitary Transformation}
 
Let us consider the following unitary transformation
\begin{equation}
R(\mu,\nu,\kappa)=\exp{\{i\mu P\}}\exp{\{i\nu D\}}\exp{\{i\kappa P^2\}}
, \label{e:tr1}
\end{equation}
where $\kappa$, $\mu$, and $\nu$ depend upon $t$.  
(We shall normally not indicate this time dependence.)  When we 
apply this transformation to the $TQ$ Schr\"odinger equation 
(\ref{e:not16a}), we have 
\begin{eqnarray}
&R(\mu,\nu,\kappa)S_1R^{-1}(\mu,\nu,\kappa)R(\mu,\nu,\kappa)\Phi(x,t)=0,&
\label{e:tr4}  \\
&\tilde{S}_1\tilde{\Phi}(x,t)=0,&  \label{e:tr8}
\end{eqnarray}
where Eq. (\ref{e:tr8}) follows from Eq. (\ref{e:tr4}) by the definitions
\begin{eqnarray}
&\tilde{S}_1=\exp{\{i\mu P\}}\exp{\{i\nu D\}}\exp{\{i\kappa P^2\}}S_1
\exp{\{-i\kappa P^2\}}\exp{\{-i\nu D\}}\exp{\{-i\mu P\}},& 
\label{e:tr12}  \\
&\tilde{\Phi}(x,t)=\exp{\{i\mu P\}}\exp{\{i\nu D\}}\exp{\{i\kappa P^2\}}
\Phi(x,t)= R(\mu,\nu,\kappa)\Phi(x,t).&
\label{e:tr16}
\end{eqnarray}

By using the theorem {\cite{wm2}}
\begin{equation}
\exp{\{B\}}A\exp{\{-B\}}=A+[B,A]+\lfrac{1}{2!}[B,[B,A]]+\cdots
\label{e:tr20}
\end{equation}
and the commutation relations in Eqs. (\ref{e:not12a}) through 
(\ref{e:not12c}),
the transformation of the operators can be carried out analytically.  
(See Ref. {\cite{wm1}}.) 
\begin{eqnarray}
R X R^{-1} &=& 
     e^{\nu}X+2\kappa e^{-\nu}P +e^{\nu}\mu I,\label{e:pdp2xa}  \\
R X^2 R^{-1}  & = & 
       e^{2\nu}X^2+4\kappa D +4\kappa^2e^{-2\nu}P^2 +2e^{2\nu}\mu X
       +4\kappa\mu P +e^{2\nu}\mu^2 I,  \label{e:pdp2x2a} \\
R P R^{-1} &=&  e^{-\nu}P,\label{e:pdp2pa}  \\
R P^2 R^{-1} &=&  e^{-2\nu}P^2,\label{e:pdp2p2a}  \\
R D R^{-1} &=&  D+\mu P +2\kappa e^{-2\nu}P^2,\label{e:pdp2da}  \\
R T R^{-1} &=&  
        T+\left({{d\mu}\over{dt}}+\mu{{d\nu}\over{dt}}\right)P
       +{{d\nu}\over{dt}}D+{{d\kappa}\over{dt}}e^{-2\nu}P^2.
     \label{e:pdp2ta}
\end{eqnarray}
Keeping in mind that $\kappa$, $\mu$, 
and $\nu$ are time-dependent, the Schr\"odinger operator $\tilde{S}_1$ is 
\begin{eqnarray}
\tilde{S}_1 & = & 
     -[1+\tilde{k}(t)]P^2+2T+\tilde{h}(t)D+\tilde{g}(t)P
-2\tilde{h}^{(2)}(t)X^2 -2\tilde{h}^{(1)}(t)X 
         -2\tilde{h}^{(0)}(t)I.  \label{e:tr24}
\end{eqnarray}
In Eq. (\ref{e:tr24}),  
the coefficients of the operators $P$, $D$, and $P^2$ are, respectively,
\begin{eqnarray}
& \tilde{g}(t)  =  2{{d\mu}\over{dt}}+e^{-\nu}\left[g(t)
      -4h^{(1)}(t)\kappa\right]+\mu\left[2{{d\nu}\over{dt}}
        +h(t)-8h^{(2)}(t)\kappa\right],&  \label{e:tr28a}  \\
& \tilde{h}(t)=2{{d\nu}\over{dt}}+h(t)-8h^{(2)}(t)\kappa, &
 \label{e:tr28b}  \\
& 1+\tilde{k}(t)=\left[-2{{d\kappa}\over{dt}}-2h(t)\kappa+8h^{(2)}
    (t)\kappa^2+k(t)+1\right]e^{-2\nu}. &   \label{e:tr28c} 
\end{eqnarray}
The coefficients of $X^2$, $X$, and $I$ are, respectively, 
\begin{eqnarray}
 & \tilde{h}^{(2)}(t) = h^{(2)}(t)e^{2\nu},~~~~~
\tilde{h}^{(1)}(t) = h^{(1)}(t)e^{\nu}+
2h^{(2)}(t)e^{2\nu}\mu, & \nonumber\\*[1mm]
 & \tilde{h}^{(0)}(t) = h^{(0)}(t)+h^{(1)}(t)e^{\nu}
\mu+h^{(2)}(t)e^{2\nu}\mu^2. & \label{e:tr28d}
\end{eqnarray}
Since the mapping $R(\mu,\nu,\kappa)$ is unitary, Eq. 
(\ref{e:tr8}), with $\tilde{S}_1$ given by Eq. (\ref{e:tr24}), has the 
same form as Eq. (\ref{e:not16a}).  Eqs. (\ref{e:tr28a}) to 
(\ref{e:tr28d}) give the conditions connecting the two $TQ$ 
equations, $S_1$ and $\tilde{S}_1$.   


\subsection{The Transformation $TQ\rightarrow TM$}

To transform Eq. (\ref{e:not16a}), $S_1\Phi=0$, into a 
$TM$-type equation (\ref{e:not16b}), $S_2\Theta=0$, we require 
that the coefficients $\tilde{h}$, $\tilde{g}$, and $\tilde{k}$ in Eqs. 
(\ref{e:tr28a}) through (\ref{e:tr28c}) satisfy the conditions:
\begin{equation}
\tilde{g}(t)=\tilde{h}(t)=0,~~~~~1+\tilde{k}(t)=f(t). 
\label{e:tr32}
\end{equation}
Under these circumstances, the operator $\tilde{S}_1$ in Eq. 
(\ref{e:tr24}) reduces to an $S_2$ operator such as in Eq. 
(\ref{e:not16b}).  Then Eqs. (\ref{e:tr28a}), 
(\ref{e:tr28b}), and the first equality in  (\ref{e:tr32}) imply 
that  $\mu$ and $\nu$ satisfy 
\begin{eqnarray}
2{{d\nu}\over{dt}}+h(t)-8h^{(2)}(t)\kappa & = & 
0, \label{e:tr36a}\\*[1mm] 
2{{d\mu}\over{dt}}+e^{-\nu}\left(g(t)-4h^{(1)}(t)
\kappa\right) & = & 0. \label{e:tr36b}
\end{eqnarray}
Also,  Eqs. (\ref{e:tr28c}) and the third equality in  
(\ref{e:tr32}) yield the equation
\begin{equation} 
{{d\kappa}\over{dt}}+h(t)\kappa-4h^{(2)}(t)
\kappa^2 =  \lfrac{1}{2}\left(1+k(t)\right)-\lfrac{1}{2}
f(t)e^{2\nu}.\label{e:tr36c}
\end{equation}

Eqs. (\ref{e:tr36a}) to
(\ref{e:tr36c}) are a set of coupled, first-order, nonlinear,  
ordinary differential equations for the functions $\kappa$, $\mu$, 
and $\nu$.  When solutions to these equations are obtained, one can 
calculate the functions ${f}^{(j)}$, $j=0,1,2$ in Eq. 
(\ref{e:not16b})  from 
\begin{equation}
f^{(j)}=\tilde{h}^{(j)},~~j=0,1,2. \label{e:tr36d}
\end{equation} 
Under these conditions, 
with $\tilde{\Phi}(x,t)$  given in Eq. (\ref{e:tr16}),
\begin{equation}
\Theta(x,t)=\tilde{\Phi}(x,t)= R(\mu,\nu,\kappa)\Phi(x,t) . 
\label{e:tr36e}
\end{equation}

We refer to Eqs. (\ref{e:tr28d}) through (\ref{e:tr36e}) 
as the ($TQ\rightarrow TM$)-connecting equations. 


\subsection{The Transformation $TM\rightarrow TO$}

As an aside, we note that, with the conditions 
\begin{equation}
\tilde{g}=\tilde{h}=\tilde{k}=0, \label{e:tr47a}
\end{equation}
a $TQ$ Schr\"odinger equation (\ref{e:not16a}) 
could be directly transformed 
into a $TO$ Schr\"odinger equation (\ref{e:not16c}), if the 
set of coupled nonlinear equations (\ref{e:tr36a}), (\ref{e:tr36b}), 
and (\ref{e:tr47b}) below had a solution.
\begin{equation}
{{d\kappa}\over{dt}}+h(t)\kappa-4h^{(2)}(t)
\kappa^2 =  \lfrac{1}{2}\left(1+k(t)\right)-\lfrac{1}{2}e^{2\nu}.
\label{e:tr47b}
\end{equation}  

Instead of using the above approach, we shall employ an 
alternative method involving a change in ``time'' variable to go 
from the $TM$  equation to the $TO$ equation.  Since we 
already have $TQ\rightarrow TM$, the time transformation  will 
complete the $TQ\rightarrow TM\rightarrow TO$ path.

We start by multiplying both sides of Eq. (\ref{e:not16b}) by $1/f(t)$, 
where we assume that $f(t)\ne 0$ for all $t$.  This yields
\begin{eqnarray}
&{\cal S}_2\Theta(x,t)  =  \left\{-P^2+{{2}\over{f(t)}}T 
                              -2{q}^{(2)}(t)X^2
  -2{q}^{(1)}(t)X-2{q}^{(0)}(t)I\right\}\Theta(x,t) 
               =  0,& \label{e:tr48}
\end{eqnarray}
where
\begin{eqnarray}
&{\cal S}_2={{1}\over{f(t)}}S_2,&  \label{e:tr52} \\
&{q}^{(j)}(t)={{f^{(j)}(t)}\over{f(t)}}.&  \label{e:tr49}
\end{eqnarray}
Now change to a new ``time'' variable $t'=t'(t)$.  Focusing on 
$(1/f(t))T$ in Eq. (\ref{e:tr48}) gives 
\begin{equation}
{{1}\over{f(t)}}T= -i{{1}\over{f(t)}}{{\partial}\over{\partial t}}
=-i{{1}\over{f(t)}}{{\partial t'}\over{\partial t}}
{{\partial}\over{\partial t'}}.\label{e:tr56}
\end{equation}
Setting the product
\begin{equation}
{{1}\over{f(t)}}{{\partial t'}\over{\partial t}}=1 \label{e:tr60}
\end{equation}
and solving for $t'(t)$  we find that 
\begin{equation}
t'-t_o'=\int_{t_o}^{t}ds\,f(s).\label{e:tr64}
\end{equation}
(Some time transformations may not be defined for all $t$ or $t'$. 
See  the examples below.)

Suppose that $t'(t)$ has an inverse.  Then, writing $t=t(t')$, we define 
the functions 
\begin{eqnarray}
 & \check{f}(t')=(f\circ t)(t'), & \nonumber\\*[1mm]
 & \check{q}^{(j)}(t')=(\tilde{f}^{(j)}\circ t)(t'),
~~~\check{h}^{(j)}(t')=(h^{(j)}\circ t)(t'),~~~j=0,1,2, & 
\label{e:tr68}\\
 & \check{\kappa}(t')=(\kappa\circ t)(t'),~~~\check{\nu}(t')
=(\nu\circ t)(t'),
~~~\check{\mu}(t')=(\mu\circ t)(t'). \label{e:tr69}
\end{eqnarray}
With the aid of the identities in Eq. (\ref{e:tr28d}) and Eq. 
(\ref{e:tr36d}), we express  $g^{(j)}$, $j=0,1,2$, as 
\begin{equation}
g^{(2)}(t')  =  {{\check{q}^{(2)}(t')}\over{\check{f}(t')}} =  
\check{h}^{(2)}(t')e^{2\check{\nu}(t')}{{1}\over{\check{f}(t')}},  
\label{e:tr72a}
\end{equation}
\begin{equation}
g^{(1)}(t')  =  {{\check{q}^{(1)}(t')}\over{\check{f}(t')}}
 =  \left[2\check{\mu}(t')\check{h}^{(2)}(t')e^{2\check{\nu}
(t')}+\check{h}^{(1)}(t')e^{\check{\nu}(t')}\right]{{1}
\over{\check{f}(t')}}, \label{e:tr72b}
\end{equation}
\begin{eqnarray}
g^{(0)}(t') & = & {{\check{q}^{(0)}(t')}\over{\check{f}(t')}}
\nonumber\\*[1mm]
 & = & \left[\check{h}^{(0)}(t')
+\check{h}^{(1)}(t')e^{\check{\nu}(t')}
\check{\mu}(t')+\check{h}^{(2)}(t')e^{2\check{\nu}(t')}
\check{\mu}^2(t')\right]{{1}\over{\check{f}(t')}}. 
\label{e:tr72c}
\end{eqnarray}

Using Eqs. (\ref{e:tr60}) and (\ref{e:tr72a}) through (\ref{e:tr72c}), 
we have mapped Eq. (\ref{e:tr48}) [that is, Eq. 
(\ref{e:not16b})] into a new Schr\"odinger equation having the form of a 
$TO$-equation (\ref{e:not16c}).  We refer to Eqs. (\ref{e:tr36a}) 
to (\ref{e:tr36c}), (\ref{e:tr64}) and its inverse, 
and (\ref{e:tr72a}) through (\ref{e:tr72c}) as the ($TM\rightarrow 
TO$)-connecting equations.  Also, we write the wave function,
$\Psi(x,t')$, as the composition
\begin{equation}
\Psi(x,t')=(\Theta\circ t)(x,t').\label{e:tr74}
\end{equation}

This completes the transformation of Eq. (\ref{e:not16b}) into 
Eq. (\ref{e:not16c}).  Now, 
let us look at the reverse transformations, $TO\rightarrow TM$ and 
$TM\rightarrow TQ$.


\subsection{The Transformations $TO\rightarrow TM$ and $TM\rightarrow TQ$}

In the $TO$ Schr\"odinger equation (\ref{e:not16c}), 
suppose that the $t'$-dependent functions, $g^{(j)}$, 
$j=0,1,2$, are suitably well-behaved, real functions but otherwise 
unspecified.  Furthermore, assume that $t'=t'(t)$ is any 
suitable, invertible function of $t$, and denote its inverse 
by $t=t(t')$.  With foresight, define the function $f(t)$ by 
\begin{equation}
f(t)={{\partial t'}\over{\partial t}}.\label{e:tr76}
\end{equation}
Then, changing the ``time'' variable in Eq. (\ref{e:not16c}) and 
multiplying the result by $f(t)$, we obtain
\begin{eqnarray}
& {S}_2\Theta(x,t)  =  \left\{-f(t)P^2+2T-2f^{(2)}(t)X^2-2f^{(1)}(t)X
-2f^{(0)}(t)I\right\}\Theta(x,t) =  0,&
\label{e:tr92}
\end{eqnarray}
where we have set 
\begin{equation}
f^{(j)}(t)=f(t)(g^{(j)}\circ t')(t),\label{e:tr96}
\end{equation}
for $j=0,1,2$, and 
\begin{equation}
\Theta(x,t) = (\Psi\circ t')(x,t).\label{e:tr88}
\end{equation}
Eq. (\ref{e:tr92}) is of the same form as the 
$TM$ Schr\"odinger equation (\ref{e:not16b}).  

We refer to 
Eqs. (\ref{e:tr76}), (\ref{e:tr88}), and (\ref{e:tr96}) as the 
($TO\rightarrow TM$)-connecting equations. 


In the $TM$ Schr\"odinger equation (\ref{e:not16b}), assume 
that $f$ and $f^{(j)}$, $j=0,1,2$, are suitably 
well-behaved, real functions of $t$.  We shall now determine 
a transformation $R(\mu,\nu,\kappa)$ of the type (\ref{e:tr1}) 
that will transform Eq. (\ref{e:not16b}) of the 
$TM$ class into a $TQ$ Schr\"odinger equation (\ref{e:not16a}).  

We choose $R(\mu,\nu,\kappa)$ such that  
\begin{equation}
R^{-1}(\mu,\nu,\kappa){S}_2R(\mu,\nu,\kappa)R^{-1}
(\mu,\nu,\kappa)\Theta(x,t)=0.\label{e:tr100}
\end{equation}
Substitute Eq. (\ref{e:tr1}) and the Schr\"odinger operator 
of Eq. (\ref{e:not16b}) into  Eq. (\ref{e:tr100}).  Then, using 
Eq. (\ref{e:tr20}) and the commutation relations 
(\ref{e:not12a})-(\ref{e:not12c}), we obtain the $TQ$ equation 
(\ref{e:not16a}), where 
\begin{eqnarray}
1+k(t) & = & 2\left({{d\kappa}\over{dt}}-2\kappa
{{d\nu}\over{dt}}\right)+8f^{(2)}(t)\kappa e^{-2\nu}
+f(t)e^{2\nu},\label{e:tr104a}\\*[1mm]
h(t) & = & -2{{d\nu}\over{dt}}+8f^{(2)}(t)\kappa e^{-2\nu},
\label{e:tr104b}\\*[1mm]
g(t) & = & -2{{d\mu}\over{dt}}e^{\nu}-8f^{(2)}(t)\kappa\mu
e^{-\nu}+f^{(1)}(t)\kappa e^{-\nu}, \label{e:tr104c}\\*[1mm]
h^{(2)}(t) & = & f^{(2)}(t)e^{-2\nu},
\label{e:tr104d}\\*[1mm]
h^{(1)}(t) & = & f^{(1)}(t)e^{-\nu}-2f^{(2)}(t)
\mu e^{-\nu}, \label{e:tr104e}\\*[1mm]
h^{(0)}(t) & = & f^{(0)}(t)-f^{(1)}(t)\mu 
+f^{(2)}(t)\mu^2.\label{e:tr104f} 
\end{eqnarray}

We refer to Eqs. (\ref{e:tr104a}) to (\ref{e:tr104f}) as the 
($TM\rightarrow TQ$)-connecting equations. As we might expect, 
Eqs. (\ref{e:tr104a})-(\ref{e:tr104c}) are equivalent to Eqs. 
(\ref{e:tr36a})-(\ref{e:tr36c}).  This can be seen by solving for 
the derivatives $d\nu/dt$ and $d\mu/dt$ and inverting Eqs. 
(\ref{e:tr104d}) to (\ref{e:tr104f}) for the functions $f^{(j)}$, 
$j=0,1,2,$.


\section{Examples with $f(t)=e^{-2\nu(t)}$}
\subsection{Form of Examples}

Now that the analysis of the transformations connecting 
the three Schr\"odinger equations (\ref{e:not16a}), (\ref{e:not16b}), 
and (\ref{e:not16c}) has been completed, we illustrate how the mapping 
works with two examples of the $TQ\rightarrow TM\rightarrow TO$ 
transformations.

But here, and in later work, we shall restrict 
$f(t)$ in Eq. (\ref{e:not16b}) to one particular form:
\begin{equation}
f(t)=e^{-2\nu(t)}.\label{e:tr38}
\end{equation}
In this case, the Schr\"odinger operator in Eq. (\ref{e:not16b}) becomes 
\begin{equation}
\hat{S}_2 = -e^{-2\nu}P^2+2T-2{f}^{(2)}(t)X^2 
-2{f}^{(1)}(t)X-2{f}^{(0)}(t)I.\label{e:tr40}
\end{equation}
The `hat' indicates the restriction (\ref{e:tr38}). 
The corresponding Schr\"odinger equation is
\begin{eqnarray}
&  \hat{S}_2\hat{\Theta}(x,t)  =  \left\{-e^{-2\nu}P^2+2T-2{f}^{(2)}
(t)X^2 -2{f}^{(1)}(t)X-2{f}^{(0)}(t)I\right\}\hat{\Theta}(x,t)
  =  0,&
\label{e:tr44} \\
&\hat{\Theta}(x,t)=\Theta(x,t,f(t)=e^{-2\nu(t)}).&
\label{e:tr45}
\end{eqnarray}
This equation is a special case of the class $TM$ equations 
(\ref{e:not16b}), where $f(t)$ is identified with $\exp{(-2\nu)}$.   
The two examples are actually of this form.  

With condition (\ref{e:tr38}), Eqs. (\ref{e:tr36a}) and (\ref{e:tr36b}) 
remain the same.  But, the right hand side of Eq. (\ref{e:tr36c}) is 
simplified to $\lfrac{1}{2}k(t)$, that is
\begin{equation}
{{d\kappa}\over{dt}}+h(t)\kappa-4h^{(2)}(t)
\kappa^2 =  \lfrac{1}{2}k(t),\label{e:tr46}
\end{equation}
which is a Riccati equation for $\kappa$.  The $f^{(j)}$, $j=0,1,2,$ in 
Eqs. (\ref{e:tr40}) and (\ref{e:tr44}) are still given by Eqs. 
(\ref{e:tr28d}) and (\ref{e:tr36d}), but $\kappa$ is now a solution 
of the Riccati equation (\ref{e:tr46}).

Since Eq. (\ref{e:tr46}) is a Riccati equation, we can proceed 
analytically in the examples below.  Furthermore, previously 
studied systems {\cite{ek1}}-{\cite{spk1}} are encompassed within 
the restriction (\ref{e:tr38}).

Before continuing with the examples, we need to specifically incorporate 
(\ref{e:tr38}) into  Eq. (\ref{e:tr64}) for the $TM\rightarrow TO$ 
transformation.  This becomes   
\begin{equation}
t'-t_o' = \int_{t_o}^{t}ds\,e^{-2\nu(s)}.  \label{e:ex0.1}
\end{equation}
In addition, Eqs. (\ref{e:tr72a}) through (\ref{e:tr72c}) yield 
\begin{equation}
g^{(2)}(t') = \check{q}^{(2)}(t')e^{2\check{\nu}(t')}
=  \check{h}^{(2)}(t')e^{4\check{\nu}(t')},  \label{e:ex0.2}
\end{equation}
\begin{equation}
g^{(1)}(t') =  \check{q}^{(1)}(t')e^{2\check{\nu}(t')}
 = 2\check{\mu}(t')\check{h}^{(2)}(t')e^{4\check{\nu}(t')}
+\check{h}^{(1)}(t')e^{3\check{\nu}(t')}, \label{e:ex0.3}
\end{equation}
\begin{eqnarray}
& g^{(0)}(t')  =  \check{q}^{(0)}(t')e^{2\check{\nu}(t')}
  =  \check{h}^{(0)}(t')
e^{2\check{\nu}(t')}+\check{h}^{(1)}(t')e^{3\check{\nu}(t')}
\check{\mu}(t')+\check{h}^{(2)}(t')e^{4\check{\nu}(t')}
\check{\mu}^2(t'). & \label{e:ex0.4}
\end{eqnarray}

\subsection{Example 1}

The first example is a frequently studied $TM$ equation 
\cite{ek1}-\cite{ps1} of the form 
\begin{equation}
\hat{S}_2\hat{\Theta}(x,t)  =  \left\{-e^{\Upsilon(t-t_o)}P^2
        +2T-\omega^2e^{-\Upsilon(t-t_o)}X^2\right\}
         \hat{\Theta}(x,t)  =  0,  
     \label{e:ex1.20}
\end{equation}
with associated $TM$ Hamiltonian 
\begin{equation}
\hat{H}_2=-\lfrac{1}{2}e^{\Upsilon(t-t_o)}\partial_{xx}
+\lfrac{1}{2}\omega^2e^{-\Upsilon(t-t_o)}x^2.
\label{e:ex1.24}
\end{equation}
Both $\Upsilon$ and $\omega^2$ are real constants and we also take 
$\omega^2$ to be positive.

First, we shall find a TQ equation related to the TM equation 
(\ref{e:ex1.20}) via the mapping (\ref{e:tr100}).  Comparing the 
Schr\"odinger equations (\ref{e:tr44}) and (\ref{e:ex1.20}), we 
observe that,
\begin{equation}
\nu=-\lfrac{1}{2}\Upsilon(t-t_o),~~~~f^{(2)}(t)=\lfrac{1}{2}\omega^2
e^{-\Upsilon(t-t_o)},~~~~f^{(1)}(t)=f^{(0)}(t)=0.\label{ex1cnds1}
\end{equation}
Therefore, Eqs. (\ref{e:tr104d}) to (\ref{e:tr104f}), yield
\begin{equation}
h^{(2)}(t)=\lfrac{1}{2}\omega^2,~~~~h^{(1)}(t)=-\omega^2
e^{-\frac{\Upsilon}{2}(t-t_o)}
\mu,~~~~h^{(0)}(t)=\lfrac{1}{2}\omega^2e^{-\Upsilon(t-t_o)}\mu^2.
\label{ex1cnds2}
\end{equation}
Conditions (\ref{ex1cnds2}) and Eqs. (\ref{e:tr36a}), (\ref{e:tr36b}), 
and (\ref{e:tr46}) together imply that
\begin{eqnarray}
 & h(t) = \Upsilon+4\omega^2\kappa, & \label{ex1cndnu}\\*[2mm]
 & 2\frac{\nsz{d\mu}}{\nsz{dt}}+e^{\frac{\Upsilon}{2}(t-t_o)}
g(t)+4\omega^2\mu\kappa=0, & \label{ex1cndmu}\\*[2mm]
 & \frac{\nsz{d\kappa}}{\nsz{dt}}+\Upsilon\kappa+2\omega^2\kappa^2
      =\lfrac{1}{2}k(t). & 
\label{ex1cndkap}
\end{eqnarray}
Since there are a nondenumerable number of choices for $k(t)$ and 
$g(t)$, there are an nondenumerable number of TQ equations 
that can be mapped into the TM equation (\ref{e:ex1.20}), with 
Eqs. (\ref{ex1cnds1}) through (\ref{ex1cndkap}) as the connecting 
equations.

A TQ Schr\"odinger equation with a 
{\it time-independent} Hamiltonian can be 
obtained by setting $k(t)=g(t)=0$ for all $t$.  With the initial condition, 
$R(t=t_o)=I$, the Riccati equation (\ref{ex1cndkap})  
with $k(t)=0$ yields the trivial solution $\kappa=0$, for all $t$, 
as the only solution.
Furthermore, Eq. (\ref{ex1cndmu}) with $g(t)=0$ and $\kappa=0$ implies that 
$\mu=0$, for all $t$, [subject to the initial condition $R(t=t_o)=I$].  Hence, 
$h(t)=\Upsilon$ and the mapping has the general form
\begin{equation}
R(0,\nu,0)=\exp\left[i\nu D\right],\label{ex1map}
\end{equation}
$\nu$  given in Eq. (\ref{ex1cnds1}).  
Under these conditions, the $TQ$ equation (\ref{e:not16a}) 
and Hamiltonian become
\begin{eqnarray}
& S_1\Phi(x,t)=\left\{-P^2+2T+\Upsilon D -\omega^2 X^2\right\}
\Phi(x,t)=0,&  \label{e:ex1.4}  \\
& H_1= - \lfrac{1}{2}\partial_{xx} -\lfrac{1}{4}\Upsilon\left(
-2ix\partial_x-i\right)+\lfrac{1}{2}\omega^2x^2. &
\label{e:ex1.8}
\end{eqnarray}

To find the equivalent $TO$ equation, we 
change the ``time'' variable.  From Eq. (\ref{e:ex0.1})  
\begin{equation}
t'-t_o' = {{1}\over{\Upsilon}}\left\{
\exp{\left[\Upsilon(t-t_o)\right]}-1\right\}.\label{e:ex1.28a}
\end{equation}
The inverse mapping is 
\begin{equation}
t-t_o={{1}\over{\Upsilon}}\ln{\left[1+\Upsilon
(t'-t_o')\right]},   \label{e:ex1.28b}
\end{equation}
where certain restrictions on $t'$ apply.  That is, 
if $(t-t_o)\in [0,\infty)$, 
then when $\Upsilon >0$, $(t'-t_o')$ lies in the interval $[0,+\infty)$ and 
when $\Upsilon<0$, $(t'-t_o')$ lies in the interval $[0,1/|\Upsilon|)$.

\indent From the ($TM\rightarrow TO$)-connecting equations 
(\ref{e:ex0.2}) to (\ref{e:ex0.4}), we see that
\begin{equation}
g^{(2)}(t')=\lfrac{1}{2}{{\omega^2}\over{\left[1+\Upsilon(t'-t_o')
\right]^2}},~~~~~g^{(1)}(t)=g^{(0)}(t)=0,\label{e:ex1.32}
\end{equation}
and the $TO$ Schr\"odinger equation (\ref{e:not16c}) 
and Hamiltonian are
\begin{eqnarray}
& {S}_3\Psi(x,t') = \left\{-P^2+2T'-
{{\omega^2}\over{\left[1+\Upsilon(t'-t_o')\right]^2}}
X^2\right\}\Psi(x,t')=0,& \label{e:e:ex1.36} \\  
& H_3 = -\lfrac{1}{2}\partial_{xx}+
{{\omega^2/2}\over{\left[1+\Upsilon(t'-t_o')\right]^2}}x^2.&
\label{e:ex1.40}
\end{eqnarray}

We shall discuss this example in detail elsewhere \cite{else}.

\subsection{Example 2}

The second example is also a $TM$ Schr\"odinger equation.
This time it is of the form 
\begin{equation}
\hat{S}_2\hat{\Theta}(x,t)=\left\{-\left({{t_o}\over{t}}\right)^a 
P^2+2T-\left({{t}\over{t_o}}\right)^b \omega^2X^2
\right\}\hat{\Theta}(x,t)=0,\label{e:ex2.24a}
\end{equation}
where $a$ and $b$ are real numbers.
The associated $TM$ Hamiltonian is 
\begin{equation}
\hat{H}_2=-\lfrac{1}{2}\left({{t_o}\over{t}}\right)^a\partial_{xx}+
\lfrac{1}{2}\omega^2\left({{t}\over{t_o}}\right)^bx^2, 
\label{e:ex2.32}
\end{equation}
for real values of $a$ and $b$.  For positive values of $a$ and $b$, this 
is the Hamiltonian system studied by Kim  {\cite{spk1}}.  We shall 
not consider the  $a=0$ case for which Eq. (\ref{e:ex2.24a}) 
already has TO form. 

As in Example 1, we shall find a TQ equation related to the TM equation 
(\ref{e:ex2.24a}) via the mapping (\ref{e:tr100}).  Comparing the 
Schr\"odinger equations (\ref{e:tr44}) and (\ref{e:ex2.24a}), we 
observe that,
\begin{equation}
\nu=\lfrac{a}{2}\ln\left(\frac{t}{t_o}\right),~~~~f^{(2)}(t)=
\lfrac{1}{2}\omega^2\left(\frac{t}{t_o}\right)^b,~~~~f^{(1)}(t)=
f^{(0)}(t)=0.\label{ex2cnds1}
\end{equation}
Therefore, Eqs. (\ref{e:tr104d}) to (\ref{e:tr104f}), yield
\begin{equation}
h^{(2)}(t)=\lfrac{1}{2}\omega^2\left(\frac{t}{t_o}\right)^{b-a},~~~~
h^{(1)}(t)=-\omega^2\left(\frac{t}{t_o}\right)^{b-a/2}\mu,~~~~
h^{(0)}(t)=\lfrac{1}{2}\omega^2\left(\frac{t}{t_o}\right)^{b}\mu^2.
\label{ex2cnds2}
\end{equation}
Conditions (\ref{ex2cnds2}) and Eqs. (\ref{e:tr36a}), (\ref{e:tr36b}), 
and (\ref{e:tr46}) together imply that
\begin{eqnarray}
 & h(t) = -\frac{\nsz{a}}{\nsz{t}}+4\omega^2\left(\frac{\nsz{t}}{\nsz{t_o}}
\right)^{b-a}\kappa, & 
\label{ex2cndnu}\\*[2mm]
 & 2\frac{\nsz{d\mu}}{\nsz{dt}}+\left(\frac{\nsz{t_o}}{\nsz{t}}
\right)^{a/2}g(t)+4\omega^2
\left(\frac{\nsz{t}}{\nsz{t_o}}\right)^{b-a}\mu\kappa=0, 
 & \label{ex2cndmu}\\*[2mm]
 & \frac{\nsz{d\kappa}}{\nsz{dt}}-\frac{\nsz{a}}{\nsz{t}}\kappa+2\omega^2
\left(\frac{\nsz{t}}{\nsz{t_o}}\right)^{b-a}\kappa^2=\lfrac{1}{2}k(t). & 
\label{ex2cndkap}
\end{eqnarray}
There are a nondenumerable number of TQ equations, depending 
upon $g(t)$ and $k(t)$, 
that can be mapped into the TM equation (\ref{e:ex2.24a}), with 
Eqs. (\ref{ex2cnds1}) through (\ref{ex2cndkap}) as the connecting 
equations.  To cite a specific example, if $g(t)=k(t)=0$, then the 
only solutions to Eqs. (\ref{ex2cndmu}) and (\ref{ex2cndkap}) 
consistent with the initial condition $R(t=t_o)=I$, is 
$\kappa=\mu=0$, for all $t$, and the transformation is of the form 
(\ref{ex1map}) with $\nu$ given in Eq. (\ref{ex2cnds1}).  

Since $h(t)=-a/t$, we see that the TQ Schr\"odinger equation (\ref{e:not1}) 
is
\begin{equation}
S_1\Phi(x,t) = \left\{-P^2+2T-{{a}\over{t}}D -
\omega^2\left(\frac{t}{t_o}\right)^{b-a}X^2\right\}\Phi(x,t)=0.\label{e:ex2.8}
\end{equation}
The corresponding time-dependent TQ Hamiltonian is 
\begin{equation}
H_1=-\lfrac{1}{2}\partial_{xx}+{{a}\over{2t}}\left(
-2ix\partial_x-i\right)+\lfrac{1}{2}\omega^2\left(\frac{t}{t_o}
\right)^{b-a}x^2.\label{e:ex2.12}
\end{equation}

In the ($TM\rightarrow TO$) transformation, 
we compute a new ``time'' variable by substituting Eq. 
(\ref{ex2cnds1}) for $\nu$ in (\ref{e:ex0.1}) and performing the integration. 
We recognize two separate cases: $a=1$ and $a\ne 0,1$.  Furthermore, 
we assume that $(t-t_o)\in [0,\infty)$.

When $a=1$, we find that
\begin{equation}
t'-t_o'=t_o\ln{\left({{t}\over{t_o}}\right)},\label{e:ex2.36b}
\end{equation}
where $(t'-t_o')\in [0,\infty)$. 
However, for $a\ne 0,1$, we have   
\begin{equation}
t'-t_o'={{t_o}\over{1-a}}\left[\left({{t}\over{t_o}}
\right)^{1-a}-1\right],\label{e:ex2.36c}
\end{equation}
where $(t'-t_o')\in [0,\lfrac{t_o}{a-1})$ if $a\in (1,\infty)$ and 
$(t'-t_o')\in [0,\infty)$ if $a\in (-\infty,0)\cup (0,1)$.

\indent From Eqs. (\ref{e:tr72b}) and (\ref{e:tr72c}), the functions 
$g^{(1)}(t')=g^{(0)}(t')=0$ in Eq. (\ref{e:not1}).  The 
$TO$ Schr\"odinger equation (\ref{e:not1}) becomes  
\begin{equation}
{S}_3\Psi(x,t')=\left\{-P^2+2T'-2g^{(2)}(t')X^2\right\}
\Psi(x,t')=0,\label{e:ex2.40}
\end{equation}
where Eq. (\ref{e:tr72a}) yields
\begin{equation}
g^{(2)}(t')= \left\{\begin{array}{ll}
\lfrac{1}{2}\omega^2\exp{\left[\left({{1+b}\over{t_o}}\right)
\left(t'-t_o'\right)\right]}, & \mbox{for $a=1$},\\
\lfrac{1}{2}\omega^2\left[1+\left({{1-a}\over{t_o}}\right)
\left(t'-t_o'\right)\right]^{\lfrac{a+b}{1-a}}, & 
\mbox{for $a\ne 0,1$}.\end{array}
\right.\label{e:ex2.46}
\end{equation}
The Hamiltonian then has the form
\begin{equation}
H_3=-\lfrac{1}{2}\partial_{xx}+g^{(2)}(t')x^2.
\label{e:ex2.48}
\end{equation}

It is also possible to map many $TM$ Schr\"odinger 
equations into a single $TO$ equation.  Example 2 provides an
illustration of this.  Look at the second equality in 
Eq. (\ref{e:ex2.46}) ($a\ne 0,1$).  For all $b=-a$, we have 
$g^{(2)}(t')=\lfrac{1}{2}\omega^2$.  Thus, we have a nondenumerable 
number of distinct $TM$ equations, determined by each $a$ and 
$b=-a$, that are mapped into a single $TO$ equation.  This 
$TO$ equation is independent of $a$ and $b$.   
This is a common phenomenon in Example 2.  

We shall also discuss this example in detail elsewhere \cite{else}.


\section{Summary}
We have developed a general method for transforming Schr\"odinger 
equations of class $TQ$ into the subclasses of $TM$ and $TO$ 
equations.  The transformation involves (i) a unitary mapping or (ii) a 
unitary mapping and a change in `time' variable.  This permits us 
to map (i) a $TQ$ Schr\"odinger equation into a 
$TM$ Schr\"odinger equation or (ii) into a $TO$ Schr\"odinger 
equation.  In paper II, we shall use these transformations to show 
that all these equations have isomorphic space-time symmetry algebras. 
We shall exploit this isomorphism and the known generators 
for $TO$ equations to obtain solutions for each of the 
Schr\"odinger equations in the class $TQ$ and the subclasses 
$TM$ and $TO$.  Then, we will compute displacement-operator 
coherent and squeezed states for each class and subclass. 
    

\section*{Acknowledgements}

MMN acknowledges the support of the United States Department of 
Energy.  DRT acknowledges
a grant from the Natural Sciences and Engineering Research Council 
of Canada.


\end{document}